# PERFORMANCE ANALYSIS OF PARALLEL POLLARD'S RHO FACTORING ALGORITHM

Anjan K Koundinya[1], Harish G[1], Srinath N K[1], Raghavendra G E[1], Pramod Y V[1], Sandeep R[1] and Punith Kumar G[1]

[1]Department of Computer Science and Engineering,
R V College of Engineering,
Bangalore, India
`annjank2@gmail.com`

## ABSTRACT

*Integer factorization is one of the vital algorithms discussed as a part of analysis of any black-box cipher suites where the cipher algorithm is based on number theory. The origin of the problem is from Discrete Logarithmic Problem which appears under the analysis of the cryptographic algorithms as seen by a cryptanalyst. The integer factorization algorithm poses a potential in computational science too, obtaining the factors of a very large number is challenging with a limited computing infrastructure. This paper analyses the Pollard's Rho heuristic with a varying input size to evaluate the performance under a multi-core environment and also to estimate the threshold for each computing infrastructure.*

## 1. INTRODUCTION

Cryptanalysis is vital study under cryptology that gives insight on strength of cryptographic algorithm; the base for such algorithm stands on the distinguishable properties like key size, number of rounds, chosen number and many more. Most of the encryption algorithm is based on Number theory aspects and strength of such algorithm purely depends on the size of the number. For instance, standard RSA algorithm implementation employs prime numbers which of 309 digits. Such algorithm gives a broad avenue for cryptanalysis study to be performed.

On the other hand from computation perspective, there is growing demand on the porting the legacy algorithm to state-of-the-art computation. This is to effectively use the optimal strength of the underlying computer architecture or computing infrastructure. A normal tendency of any algorithm is that they are either computation driven or data driven, for instance, a computation drive algorithm can be calculating a square root of number which is around 25 digits in size and data driven algorithm can be simple sorting algorithm sorting around 50,000 numbers. The former has to be addressed with multi-core architecture and latter has to be addressed by cluster computing methods.

Pollard's Rho Algorithm [6] is one such algorithm which would require computation driven solution that is well addressed under a multi-core architecture. As the number of digits in number increases the more number of cores are required to factorize the number. The most important application of this is with Discrete Logarithmic Problem (DLP). In DLP, given two large prime numbers p and g, public keyy is calculated as follows –

$$y = g^x mod\ p$$

Where *x* is kept private.

To calculate the *x* -

$$x = log_g y\ mod\ p$$

If a traditional technique such as the brute force is applied to find the value of x it would depend on the length of the prime factor. Hence a heuristic approach like the Pollard's Rho is applied to obtain the prime factors of *x*.

Earlier to the techniques of finding factor for a given number, the prominently employed algorithm was the Sieve of Erathosthenes [4]. Although the algorithm used an efficient data structure for performing the division by modulo N, it had serious flaws while operating with large numbers between 50 – 100 digits as memory key constraints for it. The algorithm is described below-

**Algorithm:** Sieve of Eratosthenes
**Input:** An integer n
**Output:** Returns an array of all prime numbers <= n

1. a[1] ← 0
2. for i ← 2 to n do
3. a[i] := 1
4. p ← 2
5. while $p_2$< n do
6. j ← $p_2$
7. while (j < n) do
8. a[j] ← 0
9. j ← j+p
10. repeat p ← p+1 until a[p] = 1
11. return(a)

The paper is organised in discuss more flavour of Pollard's Rho with respect to the required hardware implementation. The section -2 discusses in depth of the description of Pollard's Rho algorithm. The section -3 of the paper discusses about the design and implementation of the algorithm in multi-core environment. The section -4 discuss about the variety of results with the implementation.

## 2. POLLARD'S RHO ALGORITHM

Pollard's Rho algorithm [6,8] is a heuristic, i.e., the runningtime of this algorithm cannot berigorously analysed. It is saidto work very quickly when the number to be factorized hassmall factors, i.e., typically of the size of 10-12 digits. It isalso very parallelizable [3]. This is, hence, our choice ofalgorithm for implementation.Algorithms such as trial division and sieve of Eratosthenestake a lot of time because they use the brute-force approachand are only suitable for finding factors of size 3-5 digits.The other integer factorization algorithms such as the General number field sieve and its open-source implementations (MSIEVE), compute factors of any size.

158

But thesealgorithms are preferred when it is known that the numberto be factorized has large integers. This is because these algorithms take the same amount of time to find either largeor small factors [1].

**Description of Pollard's Rho Algorithm:**
**Algorithm:** Pollard's Rho
**Input:** An integer n to be factorized, and a pseudo-randomfunction f modulo n
**Output:** Factors of n

1. i ← 1
2. $x_1$ ← f(0, n-1) while true
3. do i ← i + 1
4. $x_i$ f(0, n - 1)
5. d ← gcd( | $x_{i+1}$- $x_i$ |, n)
6. if d = 1 or d = n Output d

The greatest common divisor (GCD) in Line 5 of the Algorithm Pollard's Rho gives the greatest common divisor oftwo integers which have been passed as parameters. According to the birthday paradox [5], two numbers x and y(and hence |x-y|) are congruent modulo p with probability 0.5 after($\theta\ \overline{p}$)numbers have been randomly chosen.

Ifp is a factor of n, then$p \leq \gcd(x - y, n) \leq n$. Whenthe sequence of xi start repeating after some iterations, thisis detected using the Floyd's cycle detection algorithm andPollard's Rho algorithm stops computation.This algorithm may or may not produce a factor of n. Thatis, either a correct factor or no factor is produced. This is why this algorithm falls under the class of Monte-Carlomethods.

The Floyd's detection algorithm [9] is terminating condition for the Pollard's Rho Algorithm, The description of the algorithm is given below-

**Algorithm:** Floyd's Cycle Detection
**Input:** A number $x_0 \in \{0 ...... p - 1\}$
**Output:** An index i such that $x_i = x_{2i}$ where $x_j = f(x_{j-1})\ \forall j \geq 1$

1. i←1 and $y_0$←$x_0$
2. Repeat
3. $x_i = f(x_{i-1})$ and $y_i = f(f(y_{i-1}))$
4. if $x_i$ =$y_i$ then output i and stop
5. else i← i+1

## 3. DESIGN AND IMPLEMENTATION

The pseudo-random number function used in Algorithm Pollard's Rho is of the form

$$x^2 + c$$

where 'c' is some integer otherthan -2 or 0.

Changing the c values in this function changesthe numbers of iterations that the algorithm has to performto produce a factor. For some c values, a factor may neverbe produced. The parallelization scheme performed is as follows.

International Journal of Computer Science & Information Technology (IJCSIT) Vol 5, No 2, April 2013

1. Different 'c' values are used to create different random number functions.
2. The number n that is to be factorized and a random functionis given to each CPU core available on the machine wherethis program is being run.
3. All the cores try to compute afactor, but because of the different c values, each core takesdifferent amounts of time to compute a factor.
4. If one of the cores compute a factor, the computation on all theremaining cores are stopped.
5. The number n is divided bythe computed factor to create another number $n_0$.

In realistic application, input to the algorithm would deal with large integer number; for instance the size of integer number used is around 312 digits in RSA Algorithm[7]. This is beyond the storage capacity of any built-in datatypes of any programming language. Since the parallel implementation takes OpenMP as the programming paradigm, it essential to find proper format /data structure to store such numbers. Hence GMP, an OpenSource version of GNU is employed to handle such large numbers. Now the size of the number may be restricted to available resources of the computing device. This number is given to all the cores with the old c values and thecores try to compute yet another factor of n. This processcontinues until all the prime factors of the given number are obtained.

However, implementations of Pollard's algorithm on CUDA also exist [2],but such implementation are dependent on the computingenvironment and cannot be distributed in cluster/grid environments.

## 4. RESULTS

The parallel version of Pollard's algorithm was run on one of the cores of a dual coremachine, both cores of a dual core and four cores of quad core machine. The time takento produce all factors in each case was measured(Table 4.1). The code was also run usingonly a single core, using 2 cores and then using all 4 cores of a single quad core machinemeasured the times taken to produce all factors in each case (Tables 4.2, 4.3 and 6.6).5, 50 digit numbers, 5, 100 digit numbers and 5, 200 digit numbers.

To start with the five numbers of 50 digits are tested on the machines and the time required to compute the factors is captured in the table 4.1 –

| **Single Core** | **Dual Core** | **Quad Core** |
|---|---|---|
| 18.221s | 11.162s | 8.266s |
| 24.605s | 13.234s | 9.800s |
| 27.112s | 15.334s | 10.009s |
| 21.499s | 11.857s | 8.459s |
| 22.306s | 13.706s | 9.711s |

Table 4.1: Time comparison of five 50 digit numbers on different cores machines

160

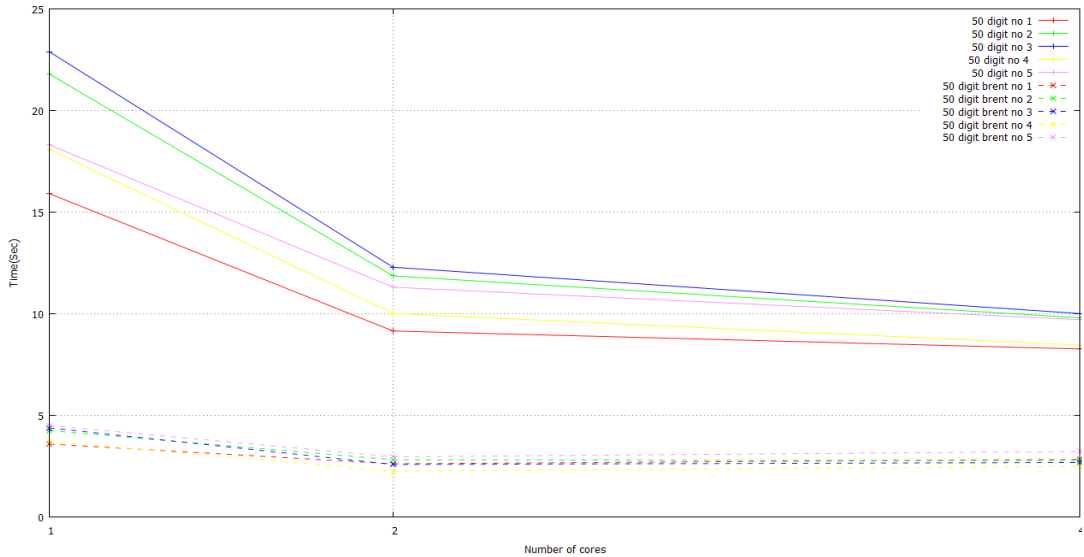

Figure 4.1: Graphical representation of 50 Digit numbers - Time Vs Cores

Testing of 100 digit numbers with different machine of variable cores, Table 4.2 captures the time taken to find factor along with graph in fig 4.2

| Single Core | Dual Core | Quad Core |
|---|---|---|
| 53.521s | 28.343s | 23.525s |
| 48.313s | 25.901s | 22.703s |
| 50.149s | 26.824s | 22.189s |
| 58.799s | 31.306s | 25.408s |
| 59.235s | 31.234s | 25.630s |

Table 4.2: Time comparison of five 100 digit numbers on different cores machines

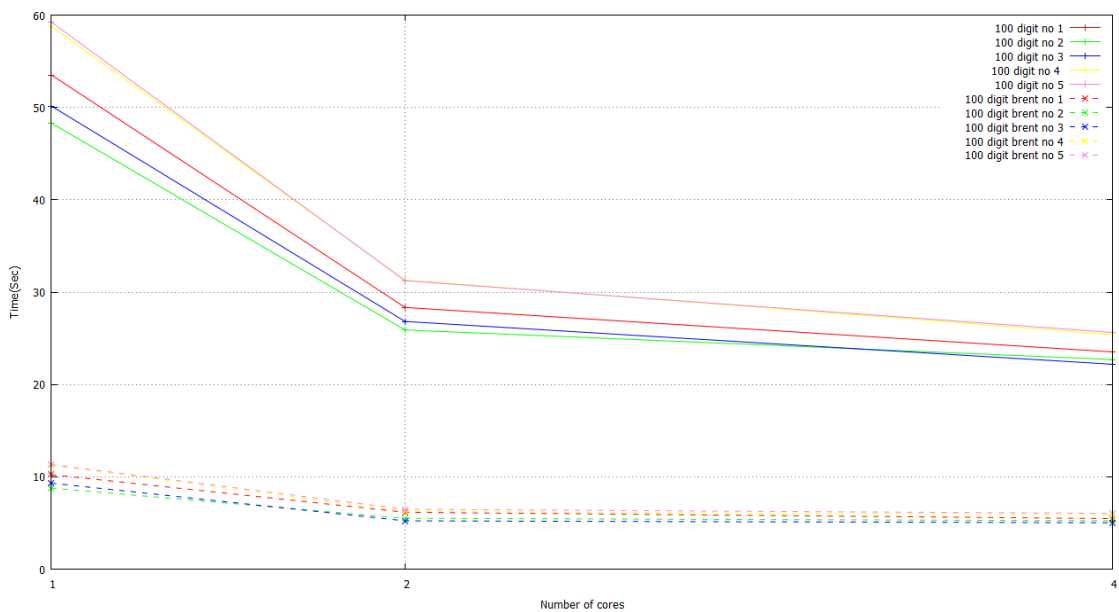

Figure 4.2: Graphical representation of 100 Digit numbers - Time Vs Cores

Further, the input number capacity is increased to 200 digits and the time required to find factor as per the our implementation is captured in table 4.3 with graphical representation in fig 4.3.

161

| Single Core | Dual Core | Quad Core |
|---|---|---|
| 137.006s | 71.104s | 52.327 |
| 136.315s | 78.461s | 64.412s |
| 268.141s | 139.403s | 113.504 |
| 146.039s | 93.481s | 77.475 |
| 117.872s | 74.880 | 63.116s |

Table 4.3: Time comparison of five 200 digit numbers on different cores machines

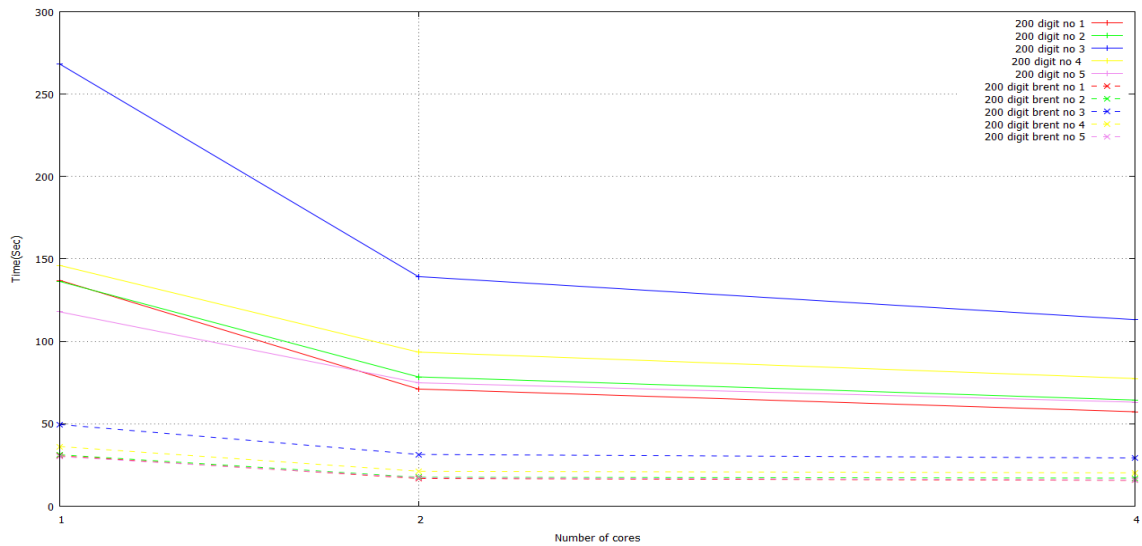

Figure 4.3: Graphical representation of five200 Digit numbers - Time Vs Cores

To conclude this section of the paper we present an average time required by each machine to find factors when the number of digits is scaled. This is graphically captured in the fig 4.4.

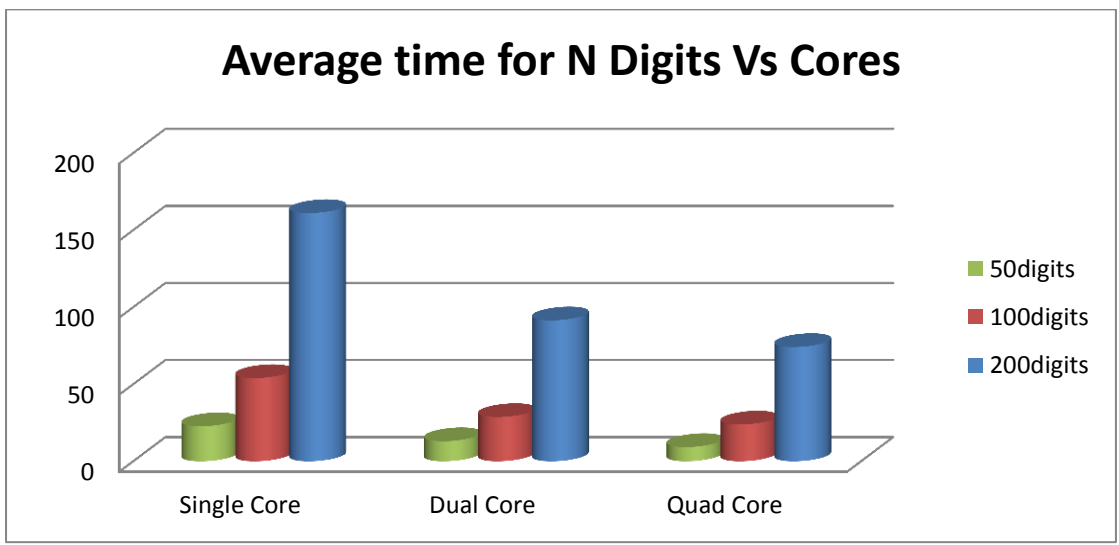

Figure 4.4: Graphical representation of Average Time Vs N-Digits Vs Cores

## 5. CONCLUSIONS

This paper presents one of the novel methods of parallelizing Pollard's Rho integer factorization and presents a coarse-grained parallelization in handling factorization computation and is based on the assumption and facts stated byBrent. This method of looking at parallel algorithm in general and Pollard's Rho algorithm in specific would givespeedup of approximately three times when the number ofcores are increased two folds. Every legacy algorithm has unique way to parallelize and make them suit in the parallelenvironment.The art of parallelizing is not concrete and is dependent on the computing environment. Due to this aspect the comparison with other legacy algorithm in parallel version may mislead the results and the change the direction ofresearch.Therefore there is immediate need to modify legacy algorithms to suit advanced computer architectures.

## ACKNOWLEDGEMENTS

Prof. Anjan K would like to thanks Late Dr. V.K Ananthashayana, Erstwhile Head, Department of Computer Science and Engineering, M.S.Ramaiah Institute of Technology, Bangalore, for igniting the passion for research. Authors would also like to thank Late. Prof. Harish G, Assistant Professor, Dept. of Computer Science and Engineering, R V College of Engineering for his consistent effort and contribution to this paper and is dedicated in remembrance of his consistent mentoring of this project.

## AUTHORS PROFILE

Anjan K Koundinya has received his B.E degree from Visveswariah Technological University, Belgaum, India in 2007 And his master degree from Department of Computer Science and Engineering, M.S. Ramaiah Institute of Technology, Bangalore, India. He has been awarded Best Performer PG 2010 and rank holder for his academic excellence. His areas of research includes Network Security and Cryptology, Adhoc Networks, Mobile Computing, Agile Software Engineering and Advanced Computing Infrastructure. He is currently working as Assistant Professor in Dept. of Computer Science and Engineering, R V College of Engineering.

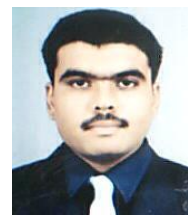

Srinath N K has his M.E degree in Systems Engineering and Operations Research from Roorkee University, in 1986 and PhD degree from Avinash Lingum University, India in 2009. His areas of research interests include Operations Research, Parallel and Distributed Computing, DBMS, Microprocessor. His is working as Professor and Head, Dept of Computer Science and Engineering, R V College of Engineering.

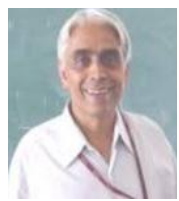